\documentclass[preprint, superscriptaddress, doublespaced, floatfix,a4paper]{revtex4-1}
\usepackage{graphicx}% Include figure files
\usepackage{siunitx}
\usepackage{float}
\usepackage{epstopdf}

\begin{document}

\title{A Hybrid Achromatic Metalens}

\author{F. Balli}
\email{fatih.balli@uky.edu}
\affiliation{University of Kentucky, Lexington, KY 40506, USA}

\author{M. Sultan}
\email{m.sultan@uky.edu}
\affiliation{University of Kentucky, Lexington, KY 40506, USA}

\author{S. Lami}
\email{sarah.lami@uky.edu}
\affiliation{University of Kentucky, Lexington, KY 40506, USA}

\author{J.T. Hastings}
\email{todd.hastings@uky.edu}
\affiliation{University of Kentucky, Lexington, KY 40506, USA}

\maketitle

\begin{figure*}
    \includegraphics[width=1\textwidth]{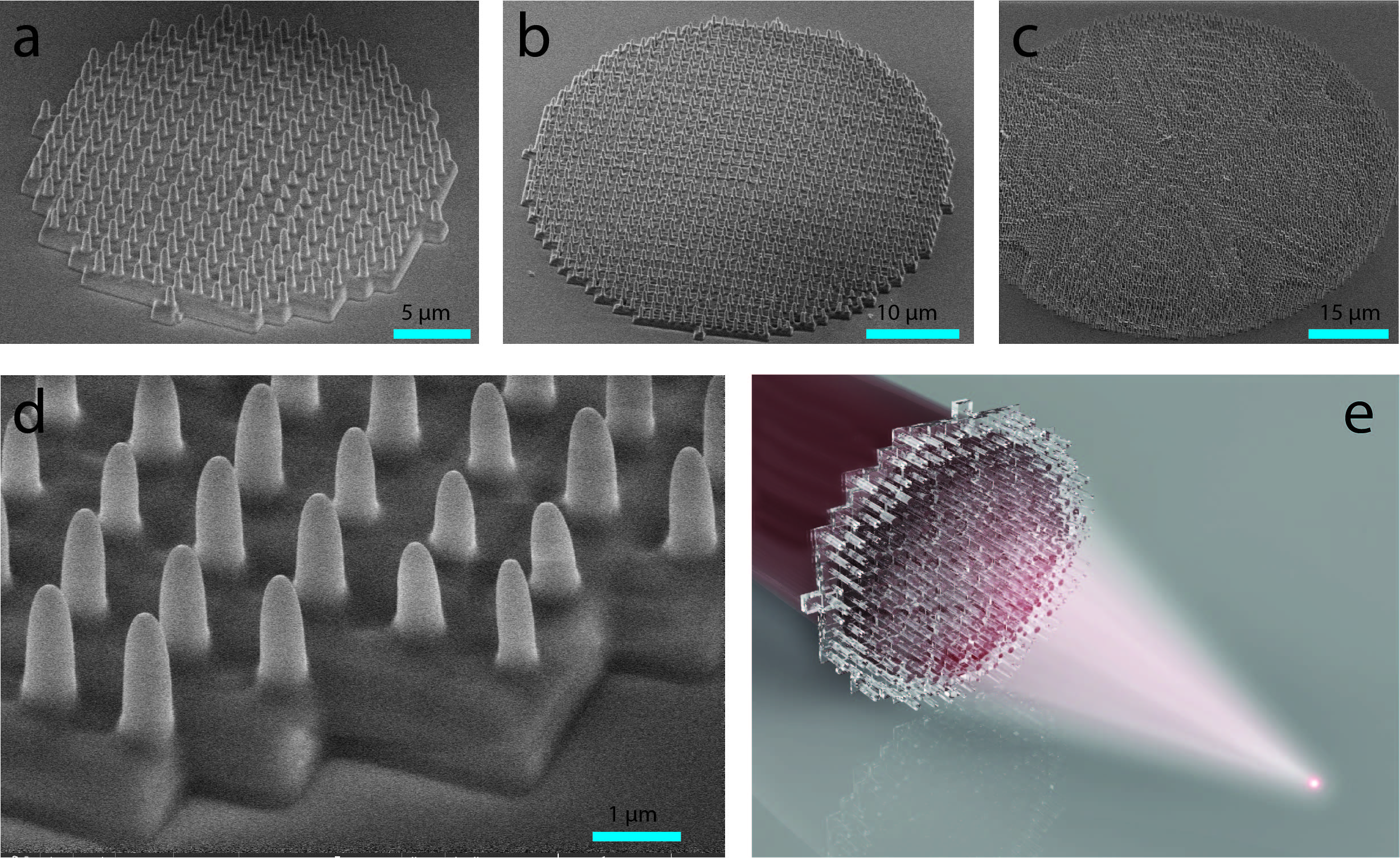} % will be changed later
    \caption{A hybrid achromatic metalens (HAML) merges a phase plate with a metalens to simultaneously correct chromatic aberration and improve focusing efficiency.  \textbf{a,b,c}, Scanning electron micrographs of HAMLs fabricated with multi-photon lithography on fused silica substrates. From left to right, aperture values are 20, 40, \SI{80}{\micro\metre}. \textbf{d}, Enlarged image of \SI{20}{\micro\metre} diameter, $0.27$ NA lens showing structure of the HAML.  Both the phase plate and the pillars of the metalens are clearly visible.  \textbf{e}, Schematic representation of a HAML illustrating broad-band focusing. }
    \label{Figure1}
\end{figure*}

\textbf{Metamaterials and metasurfaces are widely used to manipulate electromagnetic waves over a broad range of wavelengths.  Several recent efforts have focused on metalenses, ultra-thin optical elements that focus light using subwavelength structures.  Compared to their refractive counterparts, metalenses offer reduced size and weight, improved manufacturability, and new functionality such as polarization control \cite{jahaniAlldielectricMetamaterials2016, yuFlatOpticsDesigner2014,lalanne2017metalenses, yuLightPropagationPhase2011,chen2016review}.  However, metalenses that correct chromatic aberration also suffer from markedly reduced focusing efficiency.  Here we introduce a Hybrid Achromatic Metalens (HAML), shown in Fig. \ref{Figure1}, that overcomes this trade-off and offers improved focusing efficiency over a broad wavelength range from 1000 - 1800 nm.  Fabricated HAMLs demonstrated diffraction limited performance for numerical apertures (NA) of 0.27, 0.11, and 0.06 with average focusing efficiencies $\mathbf{>60\%}$ and maximum efficiencies $\mathbf{\sim80\%}$.  HAMLs can be designed by combining recursive ray-tracing and simulated phase libraries rather than computational intensive global search algorithms.  Moreover, HAMLs can be fabricated in low-refractive index materials using multi-photon lithography for customization or using molding for mass production.}  

Traditional optical components rely on the continuous phase shift of light. On the other hand, metalenses\cite{jahaniAlldielectricMetamaterials2016, yuFlatOpticsDesigner2014} are based on discontinuous phase variation \cite{lalanne2017metalenses, yuLightPropagationPhase2011}, achieved using sub-wavelength, quasi-periodic structures \cite{chen2016review}.   Metalenses are thin, light weight, can manipulate polarization\cite{lalanne2017metalenses}, and can be mass produced using planar processes similar to those used in microelectronics.  Despite these advantages, most metalenses only provide high focusing efficiency near their design wavelength as the result of chromatic aberration \cite{khorasaninejadMetalensesVisibleWavelengths2016, meinzerPlasmonicMetaatomsMetasurfaces2014,arbabi2016multiwavelength,khorasaninejadPolarizationInsensitiveMetalensesVisible2016,arbabiDielectricMetasurfacesComplete2015,arbabiMiniatureOpticalPlanar2016,ozdemirPolarizationIndependentHigh2017}.  Recent efforts have focused on creating extended libraries of nanostructure geometries, and their corresponding wavelength dependent phase shifts, that enable simultaneous correction of chromatic aberration while maintaining high transmission efficiency.\cite{groeverMetaLensDoubletVisible2017,khorasaninejadAchromaticMetalens602017,shresthaBroadbandAchromaticDielectric2018,chenBroadbandAchromaticMetalens2018,chenBroadbandAchromaticPolarizationinsensitive2019,wangBroadbandAchromaticOptical2017,wangBroadbandAchromaticMetalens2018}. However, this approach requires exotic pillar shapes with high aspect ratios that greatly increase fabrication difficulty.  Most importantly, even these complex designs have not yielded focusing efficiencies comparable to monochromatic metalenses.

Multi-level diffractive lenses (MDLs) offer an alternative approach to achromatic focusing with quasi-flat optics \cite{banerjiImagingFlatOptics2019, mohammadBroadbandImagingOne2018,wang2016chromatic}. MDLs relax the single feature-height constraint adopted by most metalenses and can be fabricated with lower resolution lithography.  These lenses are designed by global optimization algorithms that search for an appropriate surface, a technique that can be computationally intensive and limit exploration of the design space.  Moreover, focusing efficiency of MDLs still appears to be limited to values only slightly higher than achromatic metalenses.\cite{mohammadBroadbandImagingOne2018,wang2016chromatic}  In contrast, the hybrid achromatic metalenses presented here are formed by merging a phase plate and a metalens into a single thin element as shown in Fig. \ref{Figure1}.  They can be designed by combining recursive ray-tracing, similar to traditional diffractive systems, and phase libraries, similar to metalenses.  In contrast to most metalenses, HAMLs can be fabricated in low refractive-index materials.  This provides simple access to 3D geometries using multi-photon lithography or molding instead of multi-level or grayscale lithography and etching processes.  Again, relaxing feature height constraints provides additional degrees of freedom such that HAMLs can simultaneously reduce chromatic aberration, improve focusing efficiency, and preserve polarization insensitivity.  Here we design, simulate, fabricate, and characterize HAMLs with three different aperture sizes that all exhibit diffraction limited performance, improved focusing efficiency, and chromatic aberration correction over the broad spectral region from 1000 - 1800 nm. 

\begin{figure*}
    \includegraphics[width=0.9\textwidth]{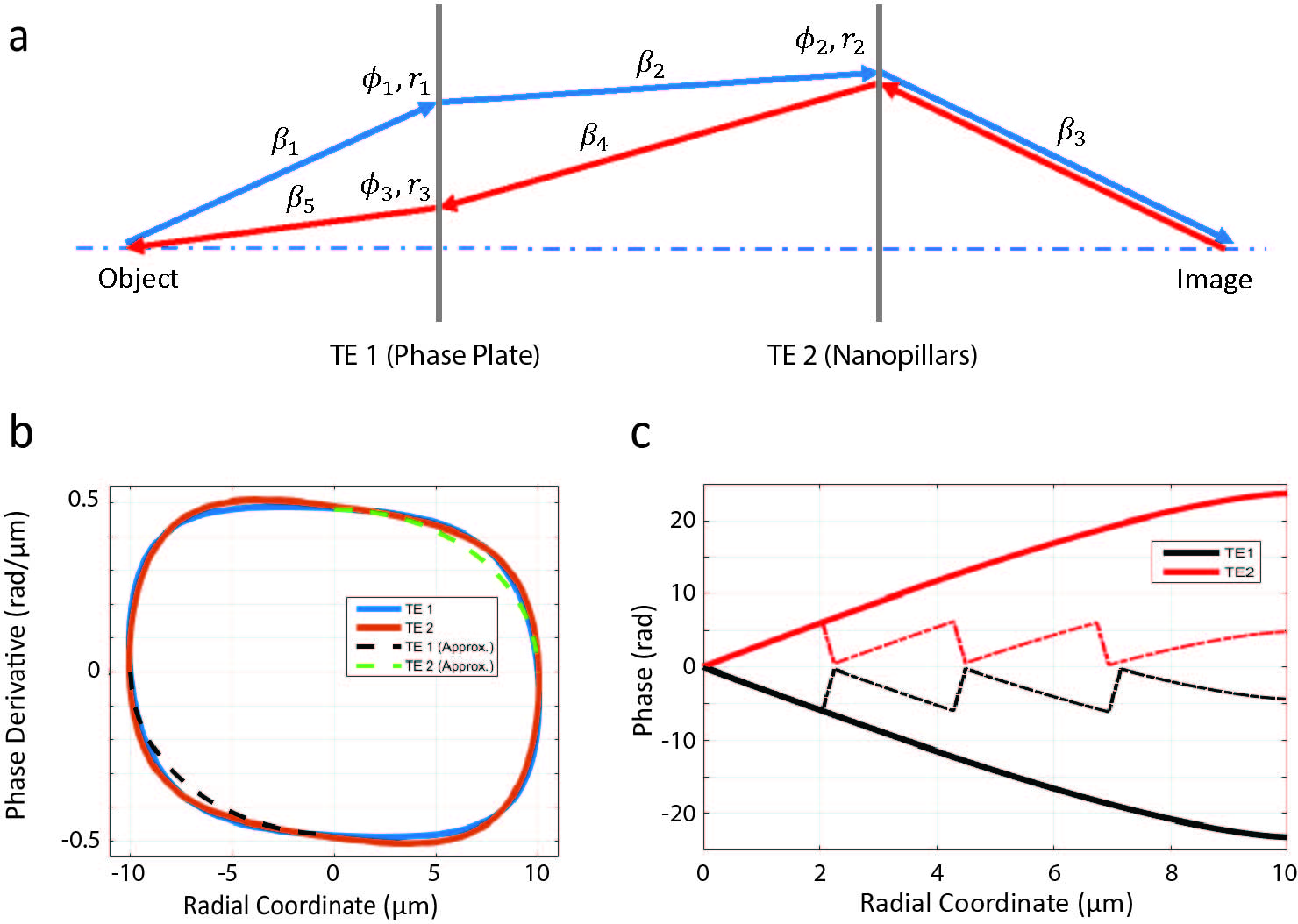}
    \caption{Design process for hybrid achromatic metalenses.  \textbf{a}, The phase derivative vs. position of the two optically thin elements (TE1: phase plate, and TE2 nanopillar metalens) are determined using recursive ray tracing. First, rays are traced with $\lambda_{min}$ from the object plane to the image plane (blue rays). Then, rays are traced with $\lambda_{max}$ from object to image (red rays).  The process repeats until the design converges. \textbf{b}, Phase derivative vs. radial coordinate for both optical elements.  Closed loops spanning both positive and negative radial coordinates indicate convergence of the algorithm. Dashed lines represent an analytic approximation of the phase derivatives for each element.  The first and second quadrants represent diverging designs, while the third and fourth quadrants provide converging designs.  \textbf{c}, Phase shift vs. radial position for each element for a \SI{20}{\micro\metre} EPD lens with 0.27 NA. Dashed lines represent the wrapped phase of the fabricated thin elements.}
    \label{Figure2}
\end{figure*}

%\section*{Theory}
Diffractive doublets can provide achromatic focusing\cite{weingartnerRealAchromaticImaging1986} and can be designed using recursive ray-tracing algorithms \cite{katoWavelengthIndependentGrating1989,farnDiffractiveDoubletCorrected1991}.  In fact, recursive ray tracing can be applied to any optically thin element (TE), including metalenses, that introduces a position dependent phase-shift.  Here we use this technique, as illustrated in Fig. \ref{Figure2}a and detailed in the methods section, to design a phase plate and a metalens which combine to correct chromatic aberration.  We then merge these two elements into a single element, a HAML, that is only a few wavelengths thick.  The algorithm first traces a ray forward from the object plane to the image plane using $\lambda_{min}$. Then, a ray is traced backward from image plane to object plane using $\lambda_{max}$.  This process repeats after replacing the phase derivative at the location of the backward ray with the phase derivative at the location of the forward ray.  The algorithm terminates when all points on the optical elements are determined and $\phi'(r)$ converges to a closed loop as shown in Fig. \ref{Figure2}b.

We find that the phase derivative as a function of radial coordinate, $r$, is well approximated by
\begin{equation}
   \phi'(r) = \phi'_{o}\sqrt{ 1-\frac{4r^2}{\textrm{EPD}^2}}
\end{equation}
where $\phi'_{o}$ and $\textrm{EPD}$ are the phase derivative at $r = 0$ $\mu m$ and entrance pupil diameter (EPD) of the lens, respectively.  This algorithm provides two possible designs for each element.  The first and second quadrants of Fig. \ref{Figure2}b yield diverging lenses, and the third and fourth quadrants yield converging lenses.  Two diverging elements will not yield a focusing lens. Two converging elements can only correct chromatic aberration if the rays cross the optical axis \cite{farnDiffractiveDoubletCorrected1991}, which is not applicable for the hybrid lens investigated here.  Combining converging and diverging designs enables correction of chromatic aberration without requiring rays to cross optical axis. We choose the third quadrant for the phase plate ($\phi'_0 = 0.47$ rad/$\mu m$) and the first quadrant for the metalens ($\phi'_0 = 0.48$ rad/$\mu m$).  The maximum error between approximate and exact phase derivatives is $0.04$ $rad/ \mu m$.  Integration over radial position gives target phase shift ,$\phi(r)$, as 
\begin{equation}
    \phi(r) = \frac{\pi \phi'_{o}}{2} \left(2r \sqrt{1-\frac{4r^2}{\textrm{EPD}^2}}+\textrm{EPD} \arcsin\left(\frac{2r}{\textrm{EPD}}\right) \right).
\end{equation}
Figure \ref{Figure2}c plots the final phase shift verses position for each element.

\begin{figure*}[!h]
    \centering
    \includegraphics[width=1\textwidth]{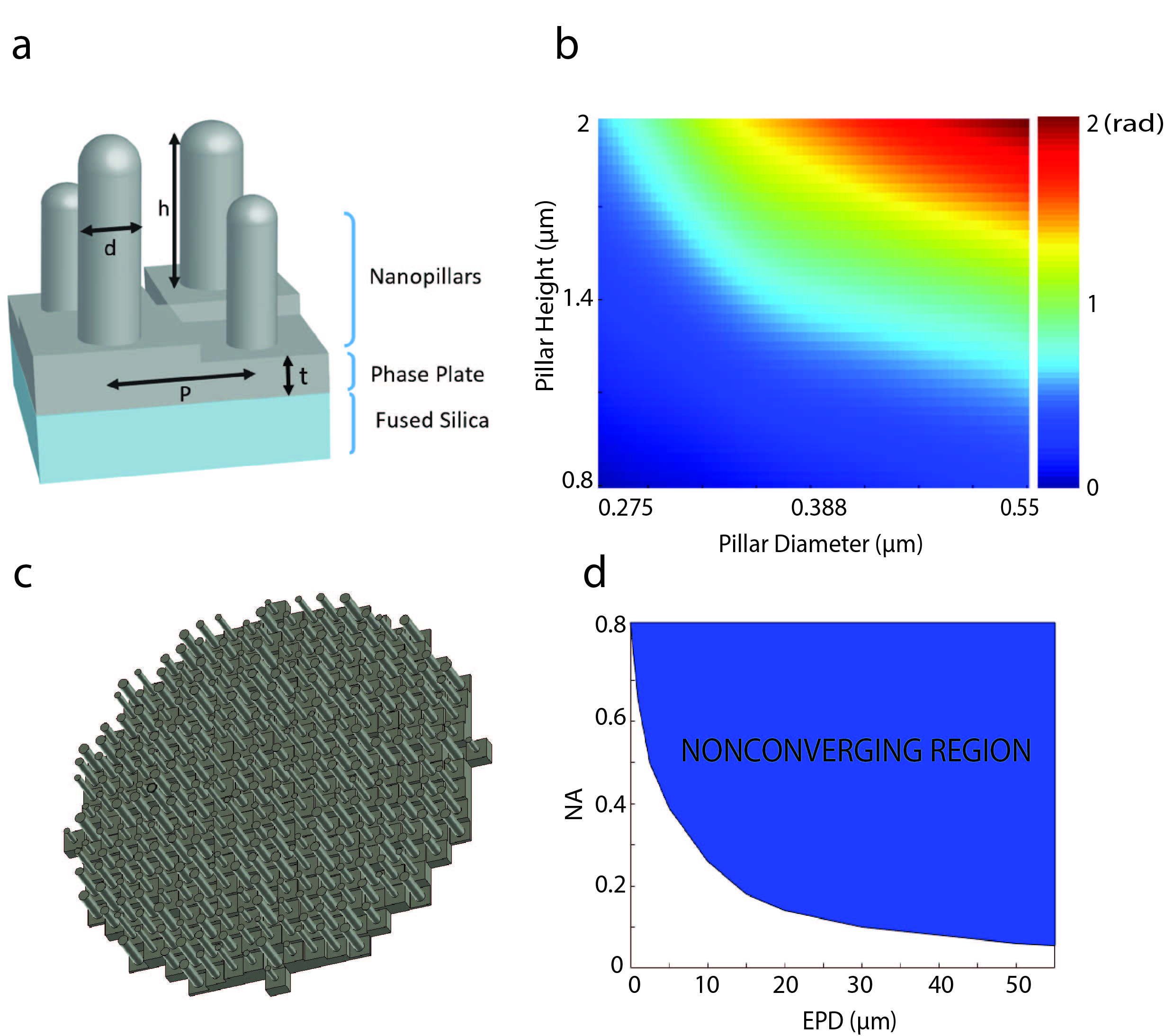}
    \caption{Hybrid achromatic metalens design.  \textbf{a,} The unit structure of a HAML consists of a phase plate and nanopillar. The thickness of the phase plate, $t$, as well as the diameter, $d$, and height, $h$, of the nanopillars are variable.  The period, $P$, is fixed.  Each set of ${t,h,d}$ yields a different phase shift.  \textbf{b,} Phase shift library for the metalens element at minimum wavelength $\lambda_1 = 1 \mu m$. Variables are diameter ($175$-$275$ $\mu m$) and height $(0.8-2$ $\mu m)$ \textbf{c,} Solid model of HAML having NA=0.27 and EPD=20 $\mu m$. \textbf{d,} Region of convergence for the recursive ray tracing algorithm (white region).  Larger combinations of NA and EPD (blue region) can still be realized experimentally and offer excellent performance.}
   \label{Figure3}
\end{figure*}

%\section*{Design}
The two building blocks of a HAML are shown in Fig. \ref{Figure3}a. The lower part contains square cross-section phase plate and the upper part contains a nanopillar-based metalens.  A similar hybrid design has been used to shape spectral transmission and produce printed color filters.\cite{limHolographicColourPrints2019}. In contrast, our hybrid metalens shapes the wavefront while maximizing broadband transmission.  We now design a phase plate and a metalens that meet the phase-shift requirements of Fig. \ref{Figure2}c at $\lambda_{min} = 1$ $\mu m$.  These will be combined into a HAML with the features shown in Fig. \ref{Figure3}a.  

The phase shift of the phase plate is given by $\phi_{pp}=\frac{2 \pi}{\lambda}(n-1)t$ where $n$ and $t$ are refractive index at $\lambda_{min}$ and thickness of the phase plate, respectively.  For the multilevel metalens, the nanopillar diameter, $d$, and height, $h$ are variable and both influence the phase shift.  The center-to-center distance between nanopillars or period, $P$, is kept fixed at $1$ $\mu m$. The period is chosen to eliminate loss due to diffraction while satisfying fabrication limitations.  This period easily meets the Nyquist sampling criterion $P < \frac{\lambda}{2NA}$. We constrain the edge-to-edge distance between pillars to reduce coupling effects\cite{lalanne2017metalenses} and avoid collapse during fabrication.  

Finite-difference time-domain (FDTD) simulations were used to build the phase library for each pillar shape which is shown in Fig. \ref{Figure3}b.  The phase shift through the nanopillars was limited to $2$ rad to ensure that the required aspect ratios could be fabricated reliably. The remaining target phase shift for the metalens is compensated by appropriately thickening the phase plate under the pillar.  This transfer of phase shift between the elements proves successful because of the lack of physical separation in the hybrid design.
 
Although there is no physical separation between thin elements, an effective distance is established by the finite size of the elements. We assume an effective thickness of $\lambda_{min}$ and check this assumption with FDTD simulations. $\lambda_{min}$ is a relatively small distance compared to the BFL and EPD, which prevents the algorithm from converging to a physically realizable design for certain combinations of NA and EPD. Figure \ref{Figure3}d shows accessible achromatic designs using the recursive algorithm.  However, through simulation, we found that an extra thickness in hyperbolic form could be added to the phase plate as a function of radial position as $t(r) = \frac{1}{ n-1}\left(\sqrt{f^2+r^2}-f \right)$.   This additional thickness enabled us to obtain higher NA values for a fixed aperture size without greatly compromising efficiency or chromatic correction.  For example, we could increase the 20 $\mu$m EPD design from  0.14 to 0.27 NA before  the efficiency began to degrade.  In all cases we limited total thickness to 3.9 $\mu$m.  Further increases in NA or EPD can be achieved by exploring new design approaches or adding a true separation between the phase plate and the metalens (increasing $d$), at the cost of a more complex fabrication process and thicker lens.  

%\section*{Experimental Results}

 To test our HAML designs, we fabricated 3 different lenses having EPD $20$, $40$ and $80$ $\mu m$ and NA values $0.27$, $0.11$ and $0.06$, respectively. The lenses were fabricated on 0.7 mm thick fused-silica substrates using``dip-in'' multi-photon lithography.  Multi-photon lithography exploits two-photon crosslinking in the focal volume of an ultrafast laser pulse to create true 3D structures in a single process step.\cite{maruo_three-dimensional_1997,vonFreymannGeorg2010TNfP}  Fabricated metalenses are shown in Fig. \ref{Figure1}a.  Lenses were characterized using a collimated beam from a supercontinuum source.  The wavelength was swept from 1000 to 1800 nm, and we measured the intensity as a function of 3D position with respect to the lens.  
 
\begin{figure*}[!h]
     \centering
    \includegraphics[width=1\textwidth]{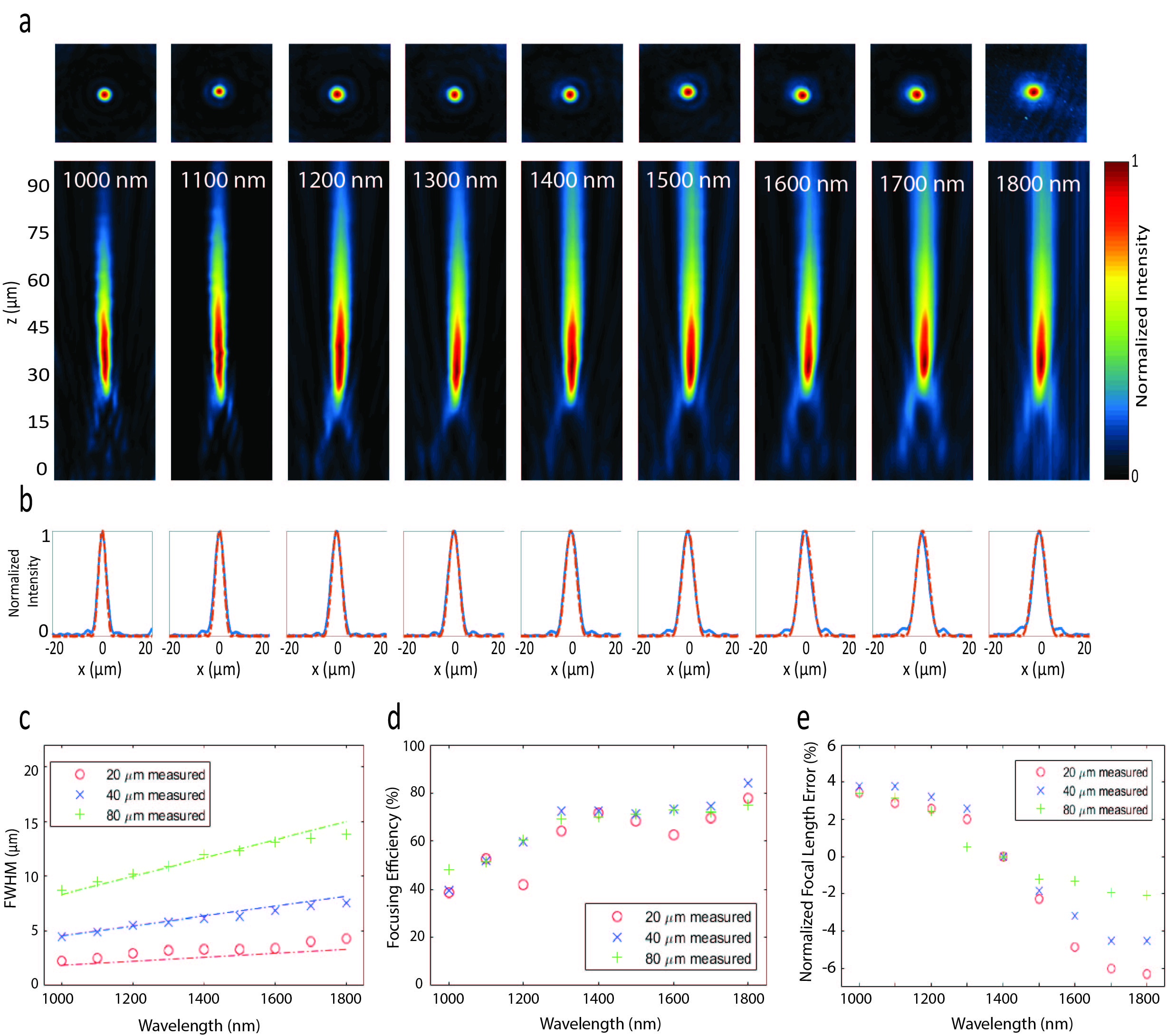}
    \caption{Experimental characterization of hybrid achromatic metalenses (HAML). \textbf{a}, Measured achromatic focusing of a collimated beam by a HAML (NA$=0.27$, EPD=\SI{20}{\micro\metre}) located at $z=0$.  Intensity distribution in focal plane (top), and intensity distribution in plane containing the optical axis (bottom).    The focal plane is $34.5$ $\mu m$ from the lens. \textbf{b}, Comparison of measured intensity (blue line) to Airy disk (dashed red line) at the focal plane.  The lens is essentially diffraction limited at all wavelengths.  \textbf{c}, Comparison between diffraction limited and measured FWHM. Red, green and blue dashed lines represent diffraction limited FWHM for $20$, $40$, \SI{80}{\micro\metre} lenses, respectively.  \textbf{d}, Focusing efficiency vs. wavelength for three different lens sizes.  \textbf{e}, Focal length error measured as a function of wavelength for three different aperture size lenses.  HAMLs improve focusing efficiency while maintaining achromatic performance across the entire wavelength range.}
    \label{Figure4}
\end{figure*}

Figure 4a plots cross sections of optical intensity in the focal plane and in a plane containing the optical axis for each wavelength.  As can be seen in Fig. \ref{Figure4}b, the lens exhibits a nearly diffraction limited point spread function at all wavelengths.  This is quantified in Fig. \ref{Figure4}c which compares measured and diffraction-limited ($\frac{0.514 \lambda}{NA}$) full-width at half-maximum (FWHM) for all three lenses. The focusing efficiency for all lenses and wavelengths is within the range $38-83\%$ as shown in Fig. \ref{Figure4}c. Focusing efficiency is defined as the ratio of total power around the on-axis focal point within the circle having radius $1.5 \times \textrm{FWHM}$ to the incident power on the metalens.\cite{banerjiImagingFlatOptics2019} Average efficiency values are $60.9\%$, $66.8\%$ and $65.6\%$ for 20, 40 and 80 $\mu$m EPD lenses, respectively.  Maximum efficiencies occur at the longest wavelength (1800 nm) and are $\sim80\%$ for all three diameters.  Finally the shift in the focal length as a function of wavelength is plotted in Fig. \ref{Figure4}e.  Here we see that the maximum shifts in focal length over the wavelength range of interest are 8.6\%, 8.2\%, and 5.4\% for 20, 40, and 80 $\mu m$ size lenses, respectively. 

%\section*{Discussion}
To put these results in context, we first compare the residual chromatic aberration of our lens vs. other planar lenses. A TiO\textsubscript{2}-based metasurface operates in the visible range with a focal length error less than $9\%$ \cite{chenBroadbandAchromaticMetalens2018}. In the NIR, a metalens with $NA=0.24$ and EPD=100 $\mu m$ exhibited a focal length error of less than $5\%$ around mean focal length in the range $1200-1600$ nm \cite{shresthaBroadbandAchromaticDielectric2018}. Here we demonstrated a similarly small focal shift of $4.8\%$ around the mean focal length, but over a broader NIR wavelength range (1000-1800 nm), albeit with a slight reduction in either NA or EPD.  Both the HAMLs discussed here and these other metalenses maintained diffraction limited performance over their design wavelength ranges.  However, HAMLs offer an expanded range of achromatic correction in the NIR.  

The advantages of HAMLs become even more apparent when we compare their focusing efficiency to that of other achromatic metalens and multi-level diffractive lenses. In the visible wavelength range, GaN-based achromatic metalenses with NA=0.106 and EPD=50 $\mu$m offer diffraction-limited performance with an average focusing efficiency of 40\% and a maximum efficiency of 67\%.  \cite{wangBroadbandAchromaticMetalens2018}. Our most similar NIR hybrid design (NA = 0.11, EPD = 40 $\mu$m) provides an average efficiency of $67\%$ and a maximum efficiency of $82\%$.  An achromatic metalens designed for the NIR (1300-1600 $\mu$m, NA=0.24, EPD=100 $\mu$m) demonstrated a maximum efficiency of $\sim55$\% \cite{shresthaBroadbandAchromaticDielectric2018} which is significantly less than even the average efficiency for these HAMLs, but again at the expense of either a slightly lower NA or EPD.  An achromatic multilevel broadband diffractive lens for visible light with NA = 0.05 and EPD = 100 $\mu$m yielded an average efficiency value of 42\% \cite{mohammadBroadbandImagingOne2018}. Our most similar NIR HAML (NA = 0.06 and EPD = 80 $\mu$m) provided an average efficiency of $65.6\%$ with a maximum efficiency of 75\%.  Thus, HAMLs provide overall improved focusing efficiency (83\% maximum and and 67\% average) compared to prior efforts, and expand the spectral bandwidth from 300 nm to 800 nm compared to  previous NIR designs.  
%\section*{Conclusion}
The bandwidth, efficiency, and diffraction-limited focusing of the hybrid achromatic metalenses presented here indicates that embracing true 3D geometries can lead to significant performance improvements for planar optics without greatly increasing their thickness.  In addition, these structures need not be overly complex to design, fabricate, or replicate.  Specifically, the phase-plate/pillar HAMLs presented here can be designed using ray tracing and small-volume, periodic FDTD simulations, and this approach should be extensible to more sophisticated combinations of refractive, diffractive, and sub-wavelength structures.  Likewise, despite their 3D nature, the relatively simple geometry of these HAMLs could also be patterned using existing grayscale and/or multilevel lithographic processing.  Such techniques could reduce dimensions and enable visible wavelength designs even if multi-photon processes with smaller features sizes are not developed in the near future.  Finally, the non-reentrant geometry and lower refractive index polymers used in these HAMLs should allow high-volume replication by molding.  These favorable performance and manufacturing attributes indicate that hybrid metaleses are excellent candidates for achromatic focusing and collimation.  Their multi-element design approach and the increased degrees of freedom inherent in true 3D structures suggest that HAMLs may ultimately improve imaging as well.  
%%TC:ignore

\noindent \textbf{References}

\bibliographystyle{unsrt}
\bibliography{references}

\newpage

\noindent \textbf{Acknowledgements}

\noindent This work was supported by Intel Corporation.  This work was performed in part at the UK Center for Nanoscale Science and Engineering, a member of the National Nanotechnology Coordinated Infrastructure (NNCI), which is supported by the National Science Foundation (ECCS-1542164).  This work used equipment supported by National Science Foundation Grant No. CMMI-1125998.
 
\noindent \textbf{Author Contributions}

\noindent J.T.H. and F.B. contributed to the theory and design of HAML. F.B. performed FDTD simulations and data analysis. F.B., M.S. and S.L. developed the HAML fabrication process. M.S. and F.B. conducted the final fabrication and optical testing.  J.T.H. supervised the project, and all authors contributed to the manuscript.

\noindent \textbf{Competing interests}

\noindent The authors declare no competing interests.

\vspace{5mm}
\noindent \textbf{\large Methods}
\vspace{5mm}

    \noindent \textbf{Recursive Ray Tracing.}  Figure \ref{Figure2}a illustrates the recursive ray-tracing algorithm used to design a phase plate and a metalens.  The angle of a ray after diffraction, $\beta _{new}$, is determined by the phase derivative, $\frac{d \phi}{dr}$, the incident angle, ${\beta_{old}}$, and the free space wavelength $\lambda$ as
    \begin{equation}
    \sin \beta _{new}=  \sin \beta _{old} +\frac{\lambda}{2 \pi} \frac{d \phi}{dr} .
    \end{equation}
    The radial position of the ray on the second optically thin element (TE) is determined by
    \begin{equation}
        r_{new} = r_{old} + d \tan \beta
    \end{equation}
    where $r_{new}$ and $r_{old}$ are the old and new radial coordinates respectively, and $d$ is distance between the TEs. We assume radial symmetry and initialize the recursive algorithm with the desired distance between the image plane and the phase plate, the distance between TEs ($d$), the back focal length (BFL), the phase derivative at radial coordinate $r_1$ on TE 1, $\phi ' (r_1)=\frac{d \phi}{dr_1}$, and the wavelength range on interest, $(\lambda_{min}, \lambda_{max})$. First, a ray is traced forward from the object plane to the image plane with $\lambda_{min}$. Knowing the $r_1, \phi'(r_1)$ pair, we can trace the ray from TE 1 to TE 2 and then TE 2 to image plane.  This procedure determines the set $(r_2,\beta_2,\phi'_2,\beta_3)$. Second, a ray is traced backward from image plane to object plane with $\lambda_{max}$ to determine $(r_3,\beta_4,\phi'_3,\beta_5)$. This process is repeated after replacing $(\phi'_3(r),r_3)$ with $(\phi'_1,r_1)$ . Each iteration will determine two $(\phi'(r),r)$ pairs for each TE and the process will terminate when all points on the TEs have been determined. Iteration can be terminated when $\phi'(r)$ converges a closed loop. For all lenses presented here the object plane was placed at infinity and thus $\beta_1=0$. 
    
    \vspace{5mm}

    \noindent \textbf{Multi-photon Lithography.} HAMLs were fabricated using a Nanoscribe Photonic Professional GT and IP-Dip photoresist (Nanoscribe GmbH, Germany) with a 63$\times$ objective in dip-in mode. The phase plate was processed as an STL file.  The metalens dimensions were exported from Lumerical FDTD to the processing package (Describe).  The hatching and slicing distance for the phase plate was set to 100 nm. The optimal laser power for phase plate writing was 17 mW, 18 mW, and 20 mW for the \SI{20}{\micro\metre}, \SI{40}{\micro\metre}, and \SI{80}{\micro\metre} diameter lenses respectively.  The pillar structures were written as single lines using piezo mode with a settling time of 150 ms.  The stage velocity was 200 $\mu m/s$. The laser power was adjusted for each pillar based on a library of interpolated measurements between the laser power and pillar radius (see supplementary information for details). After exposure, the sample was developed with 2-Methoxy-1-methylethyl acetate (PGMEA) (Avantor Performance Materials, LLC.) for 30 minutes to remove the unpolymerized resin. The sample was directly immersed in isopropyl alcohol (Fisher Scientific) for 2 minutes to remove the developer (PGMEA). Finally, To protect the metalens array from deformation due to surface tension of the IPA evaporation, the sample was immersed immediately in  a lower surface tension liquid (Novec 7100 Engineering fluid from Sigma-Aldrich) for 30 seconds and left to dry or dried in a supercritical dryer (Leica EM CPD300).  Some lenses exhibited slight bridging of the photoresist between some pillars and/or bending of a few isolated pillars.  No obvious degradation in performance was noticed as a result of these types of defects.  

    \vspace{5mm}

    \noindent \textbf{Finite-difference Time Domain Simulations.} Finite-difference time domain (FDTD) simulations were conducted using Lumerical's FDTD solver.  We used the Cauchy's equation to model the wavelength dependent refractive index of IP-Dip photoresist (Nanoscribe Gmbh.)  Its refractive index is 1.534 at $\lambda = 1000$ nm and 1.529 at $\lambda = 1800$ nm.\cite{gissibl2017refractive}  The wavelength range of interest was 1000-1800 $\mu$m.  We used a total field-scattered field (TFSF) source with pulse length, offset and bandwidth of 6.6 fs, 19 fs and 133 THz, respectively.  We used periodic and perfectly matching layer boundary conditions for transverse and longitudinal directions, respectively. Field intensity at the focal plane was obtained by the 2D frequency domain field and power monitor. We used the longitudinal component of the Poynting vector to obtain the focusing efficiency. We also use the same type of monitor in order to evaluate the power intensity distribution between focal plane and the metalens. A non-uniform mesh was used and the element sizes were automatically determined the by the software. The mesh accuracy was chosen as a compromise between accuracy, memory requirements and simulation time.

    \vspace{5mm}
    
    \noindent \textbf{Optical Testing.} A filtered supercontinuum source (SuperK EXTREME EXW-6 from NKT Photonics) was used for illumination.  The continuous bandwidth of $1150-1800$ nm is accessible using an acousto-optic filter (SuperK Select) while discrete wavelengths of $1000$ and $1100$ nm were  obtained by combining a long-pass filter (SuperK Split) and bandpass filters (Thorlabs FB1000-10 $\&$ FB1100-10). The collimated beam is aligned to the metalens' optical axis.  The focal plane of the metalens is imaged on the camera sensor by a 40x objective (Nikon M Plan 40 NA 0.55) and a tube lens (Mitutoyo MT-1). The image is captured by broadband camera (NINOX-VS-CL-640 from Raptor Photonics).  The metalens is translated along its optical axis by a motorized stage (Thorlabs MTS25-Z8) in increments of \SI{2}{\micro\metre}. For focus error measurements the increment was reduced to \SI{0.5}{\micro\metre}.  The objective and tube lens are optimized for the visible and visible-NIR wavelegnth ranges, respectively.  This introduces chromatic aberration into the measurement apparatus that must be corrected.  To accomplish this, we imaged a pinhole while sweeping the wavelength from 1000-1800 $\mu$m.  We fit a diffraction-limited Airy function to the pinhole image to determine the plane of best focus. The focal length error and z-coordinate of power intensity graph were compensated based on this measurement.  To determine focusing efficiency, we integrate the power intensity on the focal plane within a circle having radius $1.5 \times \textrm{FWHM}$ and compare to the total power incident on the surface of the fused silica in an area equal to the aperture of the lens.

\noindent \textbf{Data availability}
The data that support the plots and findings of this paper are available from J.T.H (todd.hastings@uky.edu) upon reasonable request.

%%TC:endignore
\end{document}

\title{Supplementary information for ``A Hybrid Achromatic Metalens''}

\author{F. Balli}
\email{fatih.balli@uky.edu}
\affiliation{University of Kentucky, Lexington, KY 40506, USA}

\author{M. Sultan}
\email{m.sultan@uky.edu}
\affiliation{University of Kentucky, Lexington, KY 40506, USA}

\author{S. Lami}
\email{sarah.lami@uky.edu}
\affiliation{University of Kentucky, Lexington, KY 40506, USA}

\author{J.T. Hastings}
\email{todd.hastings@uky.edu}
\affiliation{University of Kentucky, Lexington, KY 40506, USA}

\maketitle

\makeatletter
\renewcommand{\theequation}{S\arabic{equation}}
\renewcommand{\thefigure}{S\arabic{figure}}
\renewcommand{\bibnumfmt}[1]{[S#1]}
\renewcommand{\citenumfont}[1]{S#1}

\noindent \textbf{Control of Nanopillar Diameter}
    
    To optimize the multi-photon lithography process, we measured the diameter of the nanopillars as a function of laser power. We fabricated $11$ different $10 \times 10$ nanopillar arrays with average powers ranging from $14-36$ mW.  An example is shown in Figure S1a.  Figure S1b plots average diameter as a function of laser power. We fit a linear function to find the required laser power in order to obtain the target pillar diameter.
    
    \begin{figure*}[!h]
     \centering
    \includegraphics[width=1\textwidth]{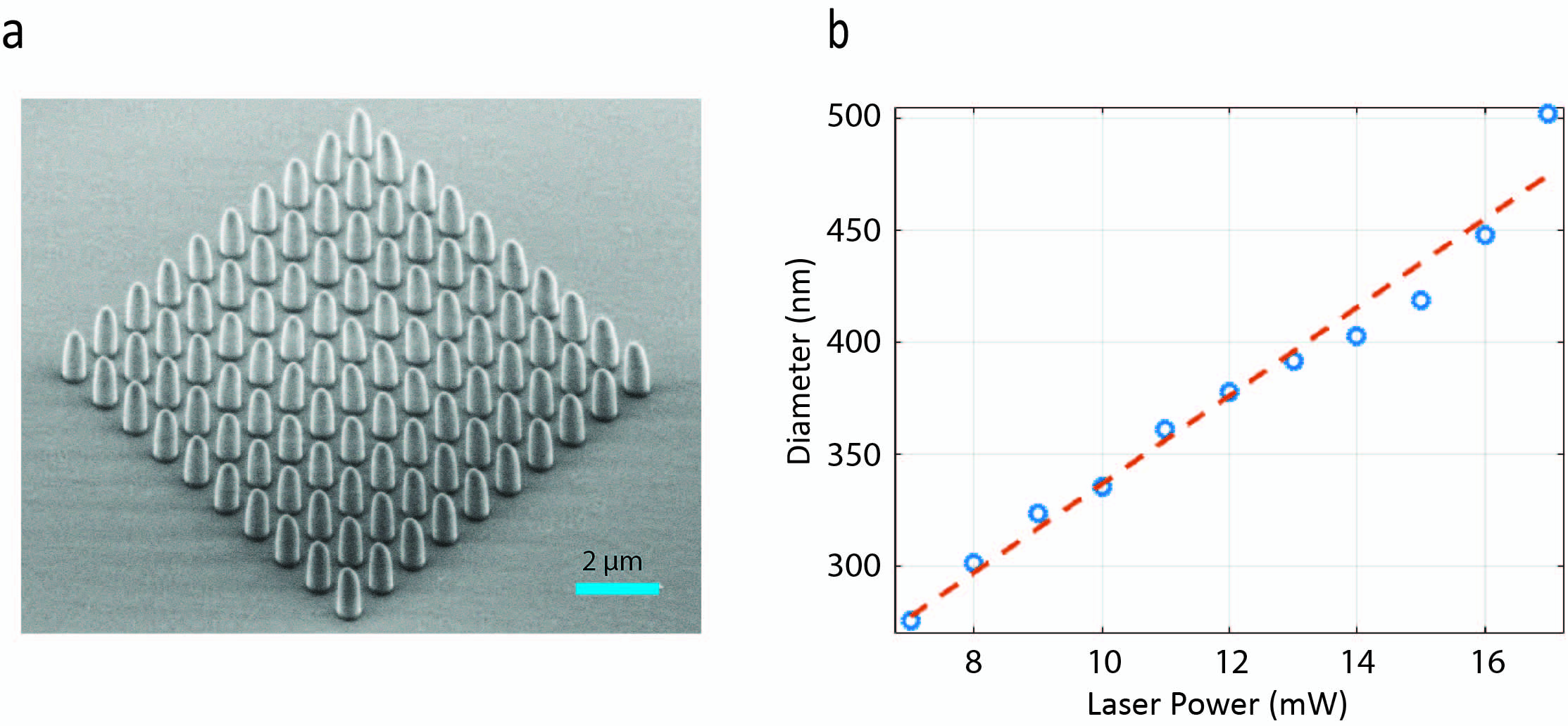}
    \caption{Determination of pillar radius vs. laser power. \textbf{a,} SEM image of one of 11 fabricated nanopillar arrays.  The laser power for this array was $13$ mW.  The height of the pillars is $1.5$ $\mu$m. \textbf{b} Pillar diameter as a function of laser power with the linear fit shown.}
     
    \end{figure*}    
    
    \noindent \textbf{Comparison of Simulated and Experimental HAML Performance}
    
\begin{figure*}[ht]
    \includegraphics[width=1\textwidth]{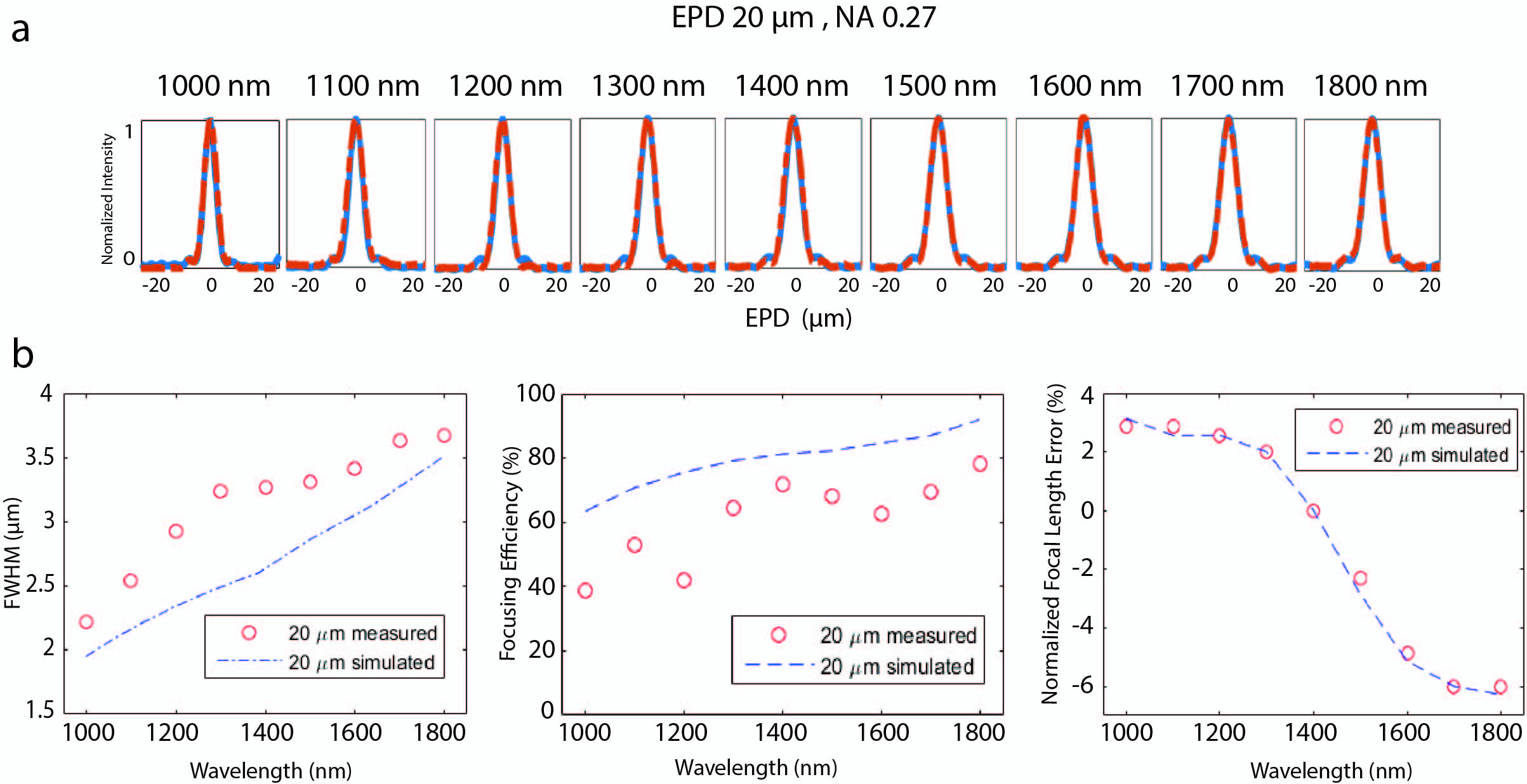}
    \caption{Comparison of simulated and experimental results for the 20 $\mu$m EPD HAML.  \textbf{a,} Simulated vs experimental focal plane intensity distributions. Red dashed line and blue continuous lines represent experimental and simulated data, respectively. \textbf{b,} Comparison of FWHM, focusing efficiency, and focusing error for experimental and simulated lenses. } 
    \label{example}
    \label{exp_vs_FDTD}
    \end{figure*}

    In the main body of the paper we compared the hybrid metalens point-spread function (PSF) to the diffraction limited PSF.  Focusing efficiency and focusing error were compared to other reported achromatic metalenses.  However, the 20 $\mu$m EPD lens is small enough for the entire lens to be simulated using FDTD.  Figure \ref{exp_vs_FDTD} compares simulated vs. experimental results for each figure of merit.  The simulated and experimental PSF match well (\ref{exp_vs_FDTD}a,b) with a maximum deviation of the FWHM at $\lambda = 1300$ nm of $0.7$ $\mu$m. The average focusing efficiency values are 60.9 \% (experimental) and 80.1 \% (simulated) as shown in Fig. \ref{exp_vs_FDTD}b.  Deviations are likely due to residual inaccuracies and defects in the fabrication.  Normalized focal length error shows excellent agreement between simulated and experimental results as well (see Fig. \ref{exp_vs_FDTD}b).
    
    \noindent \textbf{Optical Testing Apparatus}  
    
\begin{figure*}
    \includegraphics[width=1\textwidth]{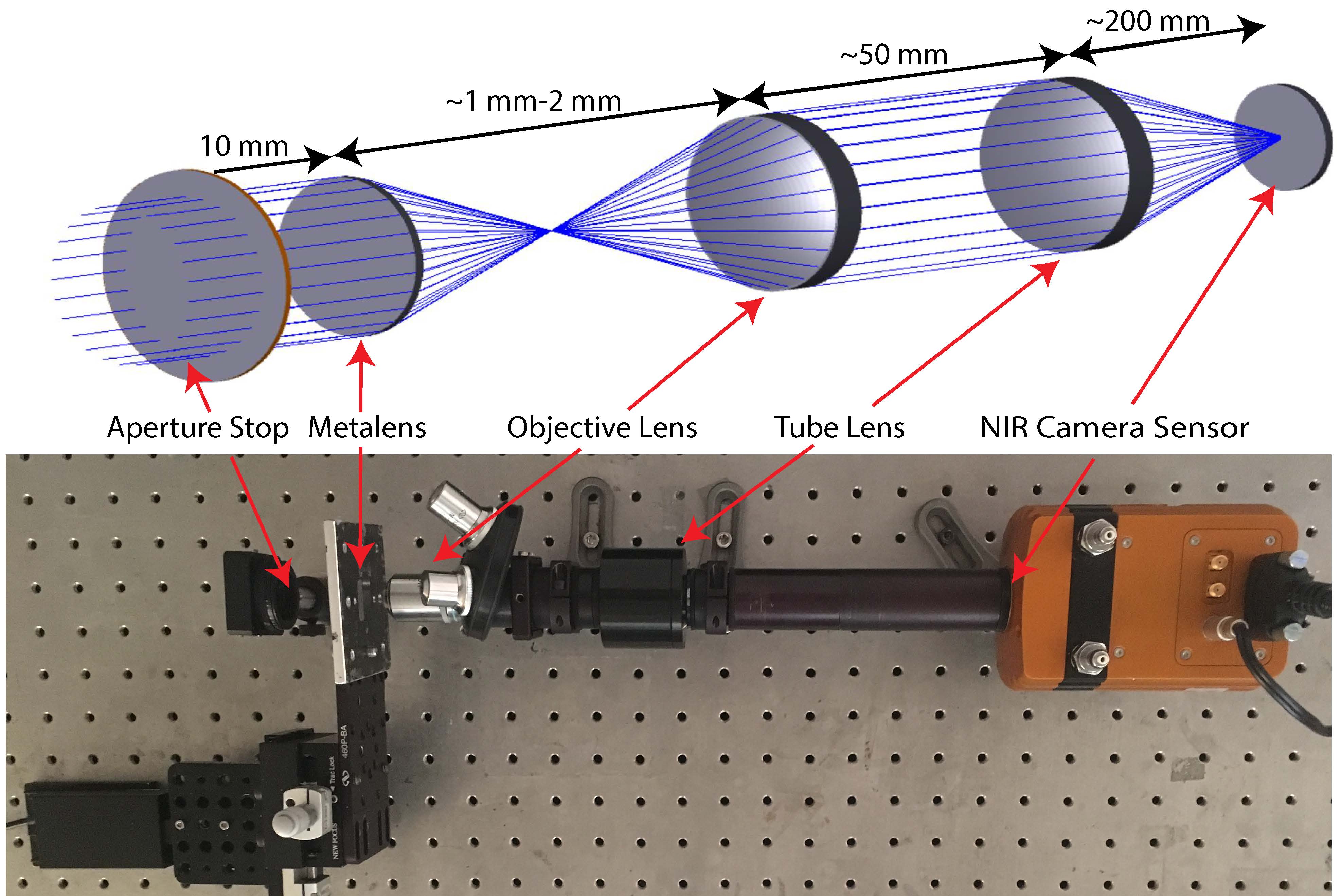}
    \caption{Optical testing setup with approximate distance between components. Distance between the metalens and objective lens is variable and controlled by the motorized translation stage. The incident light emerges from the supercontinuum light source (not shown), and the objective and tube lens relay the image formed by metalens to the focal plane array of the NIR camera with approximate 45X magnification.}
    \label{opticaltesting}
    \end{figure*}

    The apparatus for characterization of the metalenses is shown in \ref{opticaltesting}.  The key components are a supercontinuum source, a motorized translation stage, an objective lens, a tube lens, and a broadband (VIS-NIR) camera.  
    
%\section*{References}